# The Shockley-type boundary conditions for semiconductor p-n junctions at medium and high injection levels

Miron J. CRISTEA

*"Politehnica" University of Bucharest*, Bucharest, Romania

E-mail: mcris@lydo.org  Tel.: 40-214-024-915 Fax.: 40-214-300-555

**Abstract:** The classical Shockley boundary conditions are used for the determination of the minority carrier concentrations at the edges of the space-charge region of semiconductor p-n junctions. They are usually employed for the calculation of the p-n junction current/ voltage characteristic. This work demonstrates that they are valid only at low injection current levels. New analytical expressions for these boundary conditions are obtained for medium and high injection levels. These expressions are verified by calculating the current/ voltage characteristics of p-n junctions operated at high current levels.

## 1. The Shockley boundary conditions at low injection levels

In a non-degenerated semiconductor material, the concentrations of electric charge carriers are given by [1]:

$$n = N_c \cdot \exp\left(\frac{E_{fn} - E_c}{kT}\right) \quad (1)$$

$$p = N_v \cdot \exp\left(\frac{E_v - E_{fp}}{kT}\right) \quad (2)$$

with $n$ and $p$ the concentration of electrons and holes, $N_c$ and $N_v$ the densities of states in the conduction and valence bands, $E_c$ and $E_v$ the limits of the conduction and valence energy bands and $E_{fn}$ and $E_{fp}$ the quasi-Fermi energy levels in the space-charge region of the p-n junction for electrons and for holes, respectively. $k$ is the Boltzmann constant and $T$ is the absolute temperature.

The difference between the quasi-Fermi energy levels, normalized to the electric charge of the electron $q$ equals the bias voltage across the p-n junction [1]:

$$V_J = \frac{E_{fn} - E_{fp}}{q} \qquad (3)$$

Therefore, the *pn* product in the depletion region of the p-n junction is given by:

$$pn = n_i^2 \exp\left(\frac{qV_J}{kT}\right) \qquad (4)$$

with $n_i$ the intrinsic carrier concentration of the semiconductor.

Using (4), the minority carrier concentrations at the boundaries of the space-charge region are calculated:

$$p_n = \frac{n_i^2}{n_n} \exp\left(\frac{qV_J}{kT}\right) \qquad (5a)$$

$$n_p = \frac{n_i^2}{p_p} \exp\left(\frac{qV_J}{kT}\right) \qquad (5b)$$

At low injection levels, the majority carrier concentrations remain equal with the doping levels $N_D$ (donors) and $N_A$ (acceptors), therefore:

$$p_n = \frac{n_i^2}{N_D} \exp\left(\frac{qV_J}{kT}\right) \qquad (6a)$$

$$n_p = \frac{n_i^2}{N_A} \exp\left(\frac{qV_J}{kT}\right) \qquad (6b)$$

By taking into account the expressions of the minority carrier concentrations at equilibrium $p_{n0}$ and $n_{p0}$, the classical Shockley formulas are obtained [1]:

$$p_n = p_{n0} \exp\left(\frac{qV_J}{kT}\right) \qquad (7a)$$

$$n_p = n_{p0} \exp\left(\frac{qV_J}{kT}\right) \qquad (7b)$$

## 2. The case of high injection levels

At high current injection levels, both the minority and majority carrier concentrations surpass the doping level of the semiconductor, and due to the charge equilibrium condition, they are equal in magnitude. For example, in a p-n⁻ junction, where the high injection level occurs in the lightly doped n⁻ side of the junction [2]:

$$p_n = n_n \gg n_{n0} = N_D \qquad (8)$$

Using (8) and replacing $n_n$ with $p_n$ in equation (5a), the new boundary condition for the n side of the p-n junction at high injection levels is obtained:

$$p_n = n_i \exp\left(\frac{qV_J}{2kT}\right) \qquad (9a)$$

Similarly, for a p⁻-n junction at high injection levels, the boundary condition for its p⁻ side is:

$$n_p = n_i \exp\left(\frac{qV_J}{2kT}\right) \qquad (9b)$$

As verification, let's compute the current/voltage characteristic of the p-n junction at high injection levels. For example, in a p-n⁻ junction the current is given essentially by the holes injected from the p side into the low doped n⁻ side of the junction. The current equation, written for high injection level is (Appendix A):

$$J = J_p = -2qD_p\left(\frac{dp_n}{dx}\right) \qquad (10)$$

where $D_p$ is the hole diffusion constant, $J_p$ the hole current density and $J$ the total current density of the junction. Generation-recombination currents were not considered. By integrating the diffusion equation for holes with (9a) as boundary condition at the edge of the space charge region in the n⁻ side of the junction, the next formula is obtained:

$$J = \frac{2qD_p}{L_p} n_i \exp\left(\frac{qV_J}{2kT}\right) \qquad (11)$$

where $L_p$ is the diffusion length of the holes.

Three major differences arise in comparison with a classical p-n junction, at low injection levels, which has the following current/voltage characteristic [1]:

$$J = \frac{qD_p}{L_p} \frac{n_i^2}{N_D} \left[ \exp\left(\frac{qV_J}{kT}\right) - 1 \right] \quad (12)$$

(1) The diffusion constant $D_p$ appears to be doubled – due to the electric field across the n⁻ side of the junction (Appendix A);
(2) The minority concentration of holes at equilibrium $p_{n0} = n_i^2/N_D$ is replaced by the intrinsic concentration $n_i$ – due to the new boundary conditions;
(3) The voltage across the junction appears to be divided in two, a well-known fact for p-n junctions at high injection levels [2, 3], also due to the new boundary conditions.

## 3. The case of medium injection level

At the medium injection level, the simplifying conditions $n_n = n_{n0} = N_D$ - low injection level, or $n_n = p_n \gg n_{n0}$ - high injection level (pn⁻ junction), are not valid, therefore equations (5a) and (5b) must be calculated in the general case.
To do this, equation (5a) must be used in conjunction with the electric charge equilibrium equation:

$$p_n - n_n + N_D = 0 \quad (13)$$

The following expression is obtained for the minority carrier concentration $p_n$ at the boundary of the space charge region in the n side of the junction (Appendix B):

$$p_n = \left[ n_i^2 \exp\left(\frac{qV_J}{kT}\right) + \frac{N_D^2}{4} \right]^{1/2} - \frac{N_D}{2} \quad (14a)$$

Similarly, for the p side of the junction, the expression of the minority carrier concentration $n_p$ at the boundary of the space charge region is:

$$n_p = \left[ n_i^2 \exp\left(\frac{qV_J}{kT}\right) + \frac{N_A^2}{4} \right]^{1/2} - \frac{N_A}{2} \quad (14b)$$

It is easy to observe that these equations reduce to equations (9a) and (9b) when the terms including the doping levels are negligible – high injection level conditions.

They also reduce to the classical Shockley conditions in the case of low level injection, and this can be obtained by taking $N_D$ and $N_A$ respectively, as common factors in equations (14a) and (14b), expanding the square root into series and retaining only the first two terms (Appendix C).

Thus, equations (14a) and (14b) represent the general case for the boundary conditions for p-n junctions, irrespective of the injection levels at which those junctions are operated.

## 4. Conclusion

In this work the boundary conditions at the edges of the depletion region of p-n semiconductor junctions have been obtained for the general case. The analytical formulas reduce to the classical Shockley conditions in the case of low injection levels. At high injection levels, new simplified formulas have been obtained, useful for calculation of current/voltage characteristics of electronic devices employing such junctions.

## Appendix A

In the n⁻ region of the junction, the electron current density is given by:

$$J_n = q\mu_n n_n E + qD_n \frac{dn_n}{dx} \cong 0 \qquad (A1)$$

therefore drift-diffusion equilibrium is formed for electrons, as majority carriers in the n⁻ region:

$$q\mu_n n_n E = -qD_n \frac{dn_n}{dx} \qquad (A2)$$

by the means of an electric field:

$$E = -\frac{kT}{q} \frac{1}{n_n} \frac{dn_n}{dx} \qquad (A3)$$

The hole current is accordingly given by:

$$J_p = q\mu_p p_n E - qD_p \frac{dp_n}{dx} = -qD_p \left(1 + \frac{p_n}{n_n}\right) \frac{dp_n}{dx} \qquad (A4)$$

Since the n⁻ region of the junction is at high injection level ($n_n \cong p_n$), $J_p$ becomes:

$$J_p = -2qD_p \frac{dp_n}{dx} \quad (A5)$$

It's as if the built-in electric field $E$ over the n⁻ side of the junction produces the doubling of the diffusion constant of holes, in comparison with the low level injection case.

**Appendix B**

The concentration of electrons is obtained from equation (13):

$$n_n = p_n + N_D \quad (B1)$$

and then, according to (5a):

$$p_n(p_n + N_D) = n_i^2 \exp\left(\frac{qV_J}{kT}\right) \quad (B2)$$

The solution for $p_n$ from this quadratic equation is:

$$p_n = -\frac{N_D}{2} + \sqrt{\frac{N_D^2}{4} + n_i^2 \exp\left(\frac{qV_J}{kT}\right)} \quad (B3)$$

so equation (14a) is obtained.

**Appendix C**

Equation (B3) can be rewritten as:

$$p_n = \frac{N_D}{2}\left(\sqrt{1 + \frac{4n_i^2}{N_D^2}\exp\left(\frac{qV_J}{kT}\right)} - 1\right) \quad (C1)$$

In the case of low injection level, the second term of the square root is much smaller than unity; therefore the square root can be expanded into series and retained only the first two terms:

$$p_n = \frac{N_D}{2}\left(1 + \frac{2n_i^2}{N_D^2}\exp\left(\frac{qV_J}{kT}\right) - 1\right) = \frac{n_i^2}{N_D}\exp\left(\frac{qV_J}{kT}\right) \quad (C2)$$

and this is equation (6a).